\documentclass[twocolumn,prd,amsmath,superscriptaddress]{revtex4}
\usepackage{graphicx}
\usepackage{latexsym}
\usepackage{amsmath}
\usepackage{amssymb}
\usepackage{graphics}
\usepackage{color}
\usepackage{hyperref}
\usepackage{ifthen}
\usepackage{bm}
\usepackage{color}

\definecolor{darkgreen}{RGB}{0,180,0}

\newcommand\ee{\end{equation}}
\newcommand\be{\begin{equation}}
\newcommand\eea{\end{eqnarray}}
\newcommand\bea{\begin{eqnarray}}

\newcommand\mpl{M_{\rm Pl}}

\def\gsim{ \lower .75ex \hbox{$\sim$} \llap{\raise .27ex \hbox{$>$}} }

\begin{document}

\title{No-Go Theorems for Generalized Chameleon Field Theories}
\author{Junpu Wang}
\author{Lam Hui}
\affiliation{Physics Department and Institute for Strings, Cosmology and Astroparticle Physics,Columbia University, New York, NY 10027} %

\author{Justin Khoury}
\affiliation{Center for Particle Cosmology, University of Pennsylvania, Philadelphia, PA 19104
}%

\date{\today}

\begin{abstract}
The chameleon, or generalizations thereof, is a light scalar that couple to matter with gravitational strength, but whose manifestation depends on the ambient matter density. 
A key feature is that the screening mechanism suppressing its effects in high-density environments is determined by the local scalar field value.
Under very general conditions, we prove two theorems limiting its
cosmological impact: i) the Compton wavelength of such a scalar
can be at most Mpc at present cosmic density, which restricts its
impact to non-linear scales; 
ii) the conformal factor relating Einstein- and Jordan-frame scale
factors is essentially constant over the last Hubble time, which
precludes the possibility of self-acceleration.
These results imply that chameleon-like scalar fields have a negligible
effect on the linear-scale growth history; theories that invoke a
chameleon-like scalar to explain cosmic acceleration rely on
a form of dark energy rather than a genuine modified gravity effect.
Our analysis applies to a broad class of chameleon, symmetron and dilaton theories.
\end{abstract}

\maketitle

%% ADD CITE TO DEFFAYET-DGP

%%%%%%%%%%%%%%%%%%%%%%%%%%%%%%%%%%%%%%%%%%%%%%%%%%%%%%%%%%%%%%%%%%%%%%%%%%%%%%%
%%%%%%%%%%%%%%%%%%%%%%%%%%%%% SECTION: INTRODUCTION %%%%%%%%%%%%%%%%%%%%%%%%%%%
%%%%%%%%%%%%%%%%%%%%%%%%%%%%%%%%%%%%%%%%%%%%%%%%%%%%%%%%%%%%%%%%%%%%%%%%%%%%%%%

The $\Lambda$-Cold Dark Matter ($\Lambda$CDM) standard model, featuring a cosmological constant as the dark energy
component driving cosmic acceleration, will come under increased scrutiny in this decade. Upcoming large scale surveys, such
as the Dark Energy Survey, the Euclid mission, and the Large Synoptic Survey Telescope, will measure the expansion and growth histories of our
universe with unprecedented accuracy. These observations may well reveal new physics beyond $\Lambda$CDM, in the form of new
dynamical degrees of freedom in the dark sector.

It is natural to expect that such degrees of freedom (generally scalar fields) couple to both dark matter~\cite{Wetterich:1994bg,Amendola:1999er} and baryonic matter. 
Gravitationally (and universally) coupled scalars are ubiquitous in theories of high energy physics, such as string theory. To ensure consistency with local tests of gravity,
however, the effects of the scalars must be suppressed locally. This is achieved through {\it screening mechanisms}~\cite{Jain:2010ka}:
in high-density regions, such as in the solar system, the scalars develop non-linear interactions, which in turn decouple them from matter. 

There are two broad classes of screening mechanisms: 

\begin{itemize}

\item {\it Chameleon}-like mechanism, in which the scalar interactions are governed by a potential $V(\phi)$.
Whether an object is screened or not is determined by the local value of the scalar field. This
mechanism is at play in chameleon~\cite{Khoury:2003aq,Brax:2004qh,Gubser:2004uf,Mota:2006ed}, $f(R)$~\cite{Carroll:2003wy,Capozziello:2003tk,Nojiri:2003ft}, symmetron~\cite{Hinterbichler:2010es,Olive:2007aj,Pietroni:2005pv} and dilaton theories~\cite{Brax:2011ja}. These theories generally enjoy no particular symmetries, hence radiative
corrections can be important~\cite{Upadhye:2012vh}. On the other hand, since chameleons have canonical kinetic terms, there is in principle no obstruction to a UV-completion in string theory~\cite{Hinterbichler:2010wu}.

\item Vainshtein mechanism, in which the scalar non-linearities result from derivative interactions. Whether an object is screened or not
in this case is determined by derivatives of the scalar field. This
mechanism is central to the DGP model~\cite{Dvali:2000hr}
and massive gravity theories~\cite{deRham:2010kj},
which involve a scalar with the galileon symmetry
\cite{Nicolis:2008in}.
Galileon interactions are not renormalized at any order in perturbation theory~\cite{Luty:2003vm,Nicolis:2004qq,Hinterbichler:2010xn}.
On the other hand, since galileons generally propagate superluminally
on certain backgrounds, their UV completion is not a local Lorentz invariant
quantum field theory or perturbative string
theory~\cite{Adams:2006sv}.
Another possibility is a shift, but not galileon, symmetric
scalar~\cite{kmouflage}.

\end{itemize}

In this paper we focus solely on chameleon-like theories. We will prove, under very general conditions, two theorems limiting
the extent to which these theories can impact cosmological observations. The theorems apply to a broad class of these chameleon, symmetron and dilaton
theories. The key input is demanding that the Milky Way galaxy, or
the Sun, be screened, which is a necessary condition to 
satisfy local tests of gravity.

The first theorem is an upper bound on the chameleon Compton wavelength at present cosmological density:
\be
m_0^{-1} \lesssim {\rm Mpc}\,.
\label{mbound}
\ee
Since the chameleon force is Yukawa-suppressed on scales larger than $m_0^{-1}$, this implies that its effects on the large scale structure 
are restricted to non-linear scales. Any cosmological observable probing linear scales, such as redshift-space distortions,
should therefore see no deviation from general relativity in these
theories. While the bound~(\ref{mbound}) also appeared independently in~\cite{Brax:2011aw}, 
the proof presented here follows a different approach.

The second theorem pertains to the possibility of {\it self-acceleration}. Assuming that the scalar field couples universally to matter,
the theories of interest therefore involve two metrics, related by a conformal rescaling:
\be
g_{\mu\nu}^{\rm J} = A^2(\phi)g_{\mu\nu}^{\rm E}\,.
\label{confrel}
\ee
The Jordan-frame metric $g_{\mu\nu}^{\rm J}$ is the metric to which matter fields couple minimally. The Einstein-frame metric 
$g_{\mu\nu}^{\rm E}$ is by definition governed by the Einstein-Hilbert action, with constant Planck mass. Although the two frames are physically equivalent,
cosmological observations are implicitly performed in Jordan frame,
where the masses of particles are constant. Meanwhile, 
the statement that we need some form of dark energy (a component
with an equation of state $P/\rho < -1/3$) to drive
cosmic acceleration is an Einstein frame statement, where
the Friedmann equation takes its standard form.

By self-acceleration, we mean accelerated expansion in the Jordan
frame, while the Einstein-frame expansion rate is {\it not}
accelerating.
This is a sensible definition, for the lack of acceleration
in Einstein frame --- where the Einstein, and therefore the standard
Friedmann, equations hold --- is equivalent to the lack of dark energy.
In self-accelerating theories,
the observed (Jordan-frame) cosmic acceleration stems entirely
from the conformal transformation~(\ref{confrel}),
{\it i.e.}, a genuine modified gravity effect. 
The galileon provides such an example among its solutions, where
the Einstein frame metric is Minkowski and
the Jordan frame metric is de Sitter
\cite{Nicolis:2008in}.  The well-known self-accelerating
solution in the DGP model
\cite{Dvali:2000hr} can be interpreted in such a manner.

Clearly a necessary condition for self-acceleration is that the
conformal factor $A(\phi)$ varies by at least ${\cal O}(1)$ over the last
Hubble time~\cite{ftnote1}.
%\footnote{
%Relating the Jordan and Einstein frame scale factors by
%$a_J = A a_E$, it is straightforward to show
%$[a\ddot a]_J - [a \ddot a]_E = (A'/A)'$, where
%$\dot {}$ denotes derivative with respect to 
%(Jordan or Einstein frame) proper time, and ${}'$ denotes
%derivative with respect to conformal time.
%Thus, if $[a\ddot a]_E \le 0$, we must have
%$[a\ddot a]_J \le (A'/A)'$, implying $1 \lesssim \Delta A/A$
%over a (Jordan frame) Hubble time.
%}.
We will instead find for chameleon-like theories
\be
\frac{\Delta A}{A} \ll 1\,,
\label{Abound}
\ee
ruling out the possibility of self-acceleration. Jordan- and Einstein-frame metrics are indistinguishable, and cosmic acceleration
requires a negative-pressure component. 

Taken together,~(\ref{mbound}) and~(\ref{Abound}) imply that
chameleon-like scalar fields 
have a negligible effect
on density perturbations on linear scales,
and cannot account for the observed cosmic acceleration
except as some form of dark energy. 
This applies to
a broad class of chameleon, symmetron and dilaton theories,
including the popular example of $f(R)$.
In other words, 
any such model that purports to explain the observed cosmic acceleration,
and passes solar system tests,
must be doing so using some form of quintessence or vacuum energy; 
the modification of gravity has nothing to do with the acceleration
phenomenon. Nonetheless, the generalized chameleon mechanism remains interesting
as a way to hide light scalars suggested by fundamental theories.
The way to test these theories is to study small scale phenomena.
Astrophysically, chameleon scalars affect the internal dynamics~\cite{Hui:2009kc,Jain:2011ji} and stellar evolution~\cite{Chang:2010xh,Davis:2011qf,Jain:2012tn}
in dwarf galaxies in void or mildly overdense regions.

{\em Set-Up.} Consider a general scalar-tensor theory in the Einstein frame:
\be
S = \int {\rm d}^4x \sqrt{-g^{\rm E}} \left(\frac{R^{\rm E}}{16\pi G_{\rm N}} - \frac{1}{2}(\partial\phi)^2 - V(\phi)\right) + S_{\rm m}[g^{\rm J}]\,.
\ee
Matter fields described by $S_{\rm m}$ couple to $\phi$ through the
conformal factor (positive) $A(\phi)$ implicit in $g_{\mu\nu}^{\rm
  J}$. The acceleration of a {\it test particle} is influenced by the scalar:
\be
\vec{a} = -\vec{\nabla}\Phi_{\rm N} - \frac{{\rm d}\ln A(\phi)}{{\rm d}\phi} \vec{\nabla}\phi = -\vec{\nabla}\bigg(\Phi_{\rm N} + \ln A(\phi)\bigg)\,,
\label{acc}
\ee
where $\Phi_{\rm N}$ is the (Einstein frame) Newtonian
potential. The fields $\Phi_{\rm N}$ and $\phi$ obey:
\be
\vec{\nabla}^2 \Phi_{\rm N} = 4 \pi G_N A \rho \quad ; \quad
\Box \phi = V_{,\phi} + A_{,\phi} \, \rho\,,
\label{Phiphieom}
\ee
where the matter is assumed to be non-relativistic,
and $\rho$ is related to the Einstein and Jordan frame
matter densities by $\rho = \rho_{\rm E} / A = A^3 \rho_{\rm J}$ ---
defined such that $\rho$ is conserved in the usual sense
in Einstein frame~\cite{ftnote2}. 
%\footnote{
%More precisely: $U \cdot \nabla \rho = - \rho \nabla \cdot
%U$ for a pressureless fluid.
%Here, $\rho = \rho_{\rm E} / A$, $U^\mu$ is the fluid 4-velocity in Einstein
%frame ($= A \, \times$ Jordan frame velocity), and indices are contracted with the Einstein frame metric.}.
An alternative form of the $\phi$ equation of motion is
useful for comparing against the Poisson equation for $\Phi_{\rm N}$:
\be
\Box \varphi = 8 \pi G_N (V_{,\varphi} + \alpha A \rho) \, \, ; \, \,
\alpha \equiv {{\rm d} {\,\rm ln\,} A \over {\rm d}\varphi} = M_{\rm Pl}  {{\rm d}
  {\,\rm ln\,} A \over {\rm d} \phi}  \,,
\label{alphadef}
\ee
where $\varphi \equiv \phi/M_{\rm Pl}$, $M_{\rm Pl} \equiv (8\pi
G_{\rm N})^{-1/2}$, and $\alpha$ quantifies the dimensionless
scalar-matter coupling, with $\alpha \sim {\cal O}(1)$ meaning gravitational strength.

A scalar solution of interest is one where
$\phi$ takes the equilibrium value: $V_{, \phi} + A_{, \phi} \rho =
0$, {\it i.e.}, $\rho$ varies sufficiently slowly with space and time
such that gradients of $\phi$ can be neglected.
An example is cosmology, with the cosmic mean $\phi$ adiabatically tracking
the minimum $\phi_{\rm min}$ of the effective potential $V_{\rm eff} (\phi)
\equiv V(\phi) + A(\phi) \rho$ as the universe evolves.
For simplicity, we assume this minimum is unique, within the
field range of interest~\cite{ftnote3}.
%\footnote{Symmetron seems to violate this assumption, by
%having more than one minimum: different parts of
%the universe might inhabit different minima, or different domains.
%The case of the symmetron is nonetheless covered
%by our arguments, as long as one interprets the
%`field range of interest' as that within one domain.
%}.
Further, it is assumed $\phi_{\rm min}$ varies monotonically
with $\rho$, say, ${\rm d}\phi_{\rm min}/{\rm d}\rho \le 0$ ---
this is useful for implementing the idea that properties of the
scalar field vary systematically with the ambient density~\cite{ftnote4}. 
%\footnote{For instance, one would like to demand the scalar mass
%to be large, or the coupling $\alpha$ to be small, at high $\rho$.
%But our theorems do not rely on these additional assumptions.}.
Differentiating $V_{, \phi} + A_{, \phi} \rho = 0$ with respect to $\phi_{\rm min}$, it is straightforward to
show that ${\rm d}\phi_{\rm min}/{\rm d}\rho = - A_{,\phi}(\phi_{\rm min})/m^2$,
where 
\be
m^2 \equiv V_{\rm eff} {}_{,\phi\phi} (\phi_{\rm min}) =
V_{,\phi\phi} (\phi_{\rm min}) + A_{,\phi\phi} (\phi_{\rm min}) \rho
\label{mdef}
\ee
is assumed
non-negative for stability.
This means $A$ must be monotonically increasing --- hence
$V$ must be monotonically decreasing --- with $\phi$, at least over
the field range of interest.
A corollary is that $V(\phi_{\rm min}(\rho))$ and $A(\phi_{\rm min}(\rho))$ are respectively monotonically increasing and decreasing
functions of $\rho$.

We are particularly interested in the
equilibrium $\phi_{\rm min}$ at cosmic
mean density between redshifts $z=0$
and $z \simeq 1$, the period during which
the observed cosmic acceleration commences.
Let us refer to the respective equilibrium values:
$\phi_{z=0}$ and $\phi_{z\simeq 1}$. 
We are interested in theories with interesting
levels of modified gravity effects during this
period; we therefore assume:
\be
\alpha(\phi) \;\gsim\; 1 \quad {\rm for} \quad 
\phi_{z\simeq 1} \le \phi \le \phi_{z=0} \,.
\label{alphaAssume}
\ee
Note that our set-up automatically
guarantees $\phi_{z\simeq 1} \le \phi_{z=0}$.
Hence $A(\phi)$ grows with time, which is a necessary condition for self-acceleration.

{\em Generalized Screening Condition.} Consider a spherically
symmetric overdense object that is screened, meaning it sources
a scalar force that is everywhere suppressed relative to the gravitational force.
According to (\ref{acc}),
\be
\frac{{\rm d} \ln A(\phi)}{{\rm d}r} \lesssim \frac{{\rm d}\Phi_{\rm N}}{{\rm d}r}\,.
\label{screenedforce}
\ee
Both sides of the inequality are positive. The positivity of the right hand side is guaranteed by the positivity of $A\rho$;
positivity of the left will be established below.
Integrating from inside to outside the object, we have
\be
{\,\rm ln \,} \left[ 
{A(\phi_{\rm out}) \over A(\phi_{\rm in})} \right]
\lesssim \Delta \Phi_{\rm N}\,.
\ee
Here, `inside' means
the origin $r=0$; `outside' means sufficiently far out such that
$\phi_{\rm out}$ is the equilibrium value
at today's cosmic mean density: $\phi_{\rm out} = \phi_{z = 0}$.
To satisfy solar system tests, we typically demand that the Sun or the Milky Way is screened.
Both have a gravitational potential $\Phi_{\rm N} \sim -10^{-6}$, thus the screening condition is
\be
{\,\rm ln\,} \left[ {A(\phi_{z=0}) \over A(\phi_{\rm in-MW})} \right]
\lesssim 10^{-6} \, .
\label{MWscreen}
\ee
This inequality will be key in proving~(\ref{mbound}) and~(\ref{Abound}).
It makes clear that it is the gravitational potential of the object in
question, as opposed to its density alone, that ultimately determines whether it is
screened or not. 

{\em Proof of Theorems.} We first rule out self-acceleration by
proving~(\ref{Abound}). To do so requires a closer examination of the 
static and spherically symmetric equation of motion:
\be
\phi'' + \frac{2}{r} \phi' = V_{, \phi} + A_{, \phi} \rho\,, 
\label{staticeom}
\ee
where $'\equiv {\rm d}/{\rm d}r$. This is subject to the boundary conditions $\phi'|_{r=0} = 0$ and $\phi \rightarrow_{r \rightarrow \infty} \phi_{z = 0}$.
Although $\phi$ tends to its equilibrium value asymptotically, we make no such assumption at the origin, {\it i.e.}, 
$\phi|_{r=0} \equiv \phi_{\rm in}$ need {\it not} coincide with $\phi_{\rm min}( \rho_{\rm in})$. We distinguish 3 cases:

\noindent $\bullet$ {\it Case 1}: Suppose $V_{, \phi} + A_{, \phi} \rho \simeq 0$ at $r=0$, that is,
$\phi_{\rm in} \simeq \phi_{\rm min}( \rho_{\rm in})$. This is the thin-shell case of standard chameleons~\cite{Khoury:2003aq}.
Since $\rho_{\rm MW} \gg \rho_{z\simeq 1}$, our monotonicity assumptions imply $A(\phi_{z\simeq 1}) \ge A(\phi_{\rm in-MW})$, thus
\be
\ln \left[\frac{A(\phi_{z=0})}{A(\phi_{z\simeq 1})}\right] 
\le \ln \left[\frac{A(\phi_{z=0})}{A(\phi_{\rm in-MW})}\right]  
\lesssim 10^{-6}\,.
\label{Delphibound}
\ee
This proves (\ref{Abound}) in this case. 
 
\noindent $\bullet$ {\it Case 2}: Suppose $A_{, \phi}\rho \gg -V_{, \phi}$ at $r = 0$, which is the case relevant to symmetrons~\cite{Hinterbichler:2010es}.
Given our assumption that $V_{\rm eff} = V(\phi) + A(\phi) \rho $ has a unique minimum, this implies $\phi_{\rm in}  \geq  \phi_{\rm min}( \rho_{\rm in})$.
Because $\phi'|_{r=0} = 0$, it follows from~(\ref{staticeom}) that $\phi''|_{r=0} > 0$, and thus $\phi'|_{r>0} > 0$. 
And since $\phi'$ is continuous at the surface of the object, to satisfy $\phi \rightarrow_{r \rightarrow \infty} \phi_{z = 0}$ we must therefore have
$\phi_{\rm in} < \phi_{z = 0}$. In other words, Case 2 corresponds to
\be
\phi_{\rm min}( \rho_{\rm in}) \leq \phi_{\rm in} <  \phi_{z = 0}\,.
\label{case2ineq}
\ee
Unlike Case 1, $\phi_{\rm in-MW}$ is not {\it a priori} constrained to be smaller (or greater) than 
$\phi_{z\simeq 1}$. If $\phi_{z\simeq 1} \geq \phi_{\rm in-MW}$, then as in Case 1 we are led to~(\ref{Delphibound}), and self-acceleration is ruled out.
The other possibility, $\phi_{z\simeq 1} < \phi_{\rm in-MW}$, is inconsistent with screening the Milky Way. Indeed, in this case $\phi$ falls within the range~(\ref{alphaAssume}) where $\alpha(\phi) \;\gsim\; 1$,
and~(\ref{alphadef}) can be approximated by $\nabla^2 \varphi \sim 8\pi G_{\rm N} \alpha A\rho$. Comparing with the Poisson equation $\nabla^2 \Phi_{\rm N} = 4\pi G_{\rm N} A
\rho$, it is clear the resulting scalar force is not small compared to the gravitational force, thus invalidating the screening of the Milky Way.

\noindent $\bullet$ {\it Case 3}: Suppose $A_{, \phi}\rho \ll -V_{, \phi}$ at $r = 0$, that is, $\phi_{\rm in} \leq \phi_{\rm min}( \rho_{\rm in})$. In this case, all inequalities are
reversed relative to Case 2, and instead of~(\ref{case2ineq}) we conclude $\phi_{\rm min}( \rho_{\rm in}) \geq \phi_{\rm in} >  \phi_{z = 0}$. But this is inconsistent
with our assumption that $\phi_{\rm min}(\rho)$ is monotonically decreasing, hence we can ignore this case. 

To summarize, the only phenomenologically viable possibilities are Case 1 and Case 2 with $\phi_{z\simeq 1} \geq \phi_{\rm in-MW}$. In both cases we are led to~(\ref{Delphibound}).
The very small $\Delta A/A$ over cosmological time scales precludes self-acceleration.

To establish the bound~(\ref{mbound}) on $m_0^{-1}$, consider the (Einstein-frame) cosmological evolution equation:
\be
\ddot{\phi} + 3H \dot{\phi} = -V_{,\phi}  - A_{,\phi} \rho\,,
\label{phicosmo}
\ee
where $\rho$ is the total (dark matter plus baryonic) non-relativistic matter component, and $H\equiv \dot{a}_{\rm E}/a_{\rm E}$ is the Einstein-frame Hubble parameter. Since
$A(\phi)\rho\sim  H^2M_{\rm Pl}^2$ from the Friedmann equation, the density term in~(\ref{phicosmo}) exerts a significant pull on $\phi(t)$. 
The potential prevents a rapid roll-off of $\phi$ by canceling the density term to good accuracy: $V_{,\phi} \simeq - A_{,\phi} \rho$.
This cancellation must be effective over at least the last Hubble time, {\it i.e.} $\phi$ must track adiabatically the minimum of the effective potential. 
Differentiating this relation with respect to time, and using~(\ref{mdef}) together with the conservation law $\dot{\rho} = -3H\rho$ and the Friedmann relation,
we find
\be
m^2 \simeq \frac{3HA_{,\phi}\rho}{\dot{\phi}} \sim H\frac{{\rm d} t}{{\rm d}\ln A} \alpha^2(\phi) H^2 \,.
\ee
The factor of $H^{-1}{\rm d}\ln A/{\rm d}t$ is the change of $\ln A$ over the last Hubble time, which from~(\ref{Delphibound}) is less than $10^{-6}$. Thus
\be
m^2 \;\gsim\; 10^6\alpha^2(\phi) H^2 \,.
\ee
Using~(\ref{alphaAssume}), it follows that $m_0^{-1}\lesssim 10^{-3}H_0^{-1}\sim {\rm Mpc}$,
as we wanted to show.

{\em Discussion.} As with any no-go theorem, the key question is which of its
assumptions can be circumvented? One option is to relax the assumption of adiabatic
tracking for the cosmological scalar field. While this opens up a wider range of possibilities,
a generic outcome is that $\phi$ undergoes large field excursions on cosmological time scales.
Indeed, if the two terms on the right-hand side of (\ref{phicosmo}) do not (approximately) cancel each other,
the resulting acceleration $\vert\ddot{\phi}\vert \sim \mpl H^2$ --- either positive, if $V$ dominates,
or negative, if $A\rho$ dominates --- will drive the field by an amount $\vert\Delta \phi \vert\sim \mpl (\Delta z)^2$
over a redshift difference $\Delta z$. In particular, $\vert\Delta \phi \vert\sim \mpl$ for $\Delta z \simeq 1$.
It is unclear whether such large field excursions are consistent with the Milky Way being screened.

%For possibility (1), if adiabaticity is violated for too large a $\Delta z$, then we risk getting into the regime 
%$\phi_{z=0} \ge  \phi_{\rm in-MW} >  \phi_{\Delta z}$, which from our discussion in Case 2 above is inconsistent
%with the Milky Way being screened. Thus the most we can tolerate is $\phi_{z=0}\ge \phi_{\rm in-MW} \simeq  \phi_{\Delta z}$;
%given~(\ref{MWscreen}) and $\alpha(\phi)\gtrsim O(1)$, this requires $\Delta z \lesssim 10^{-3}$, which is fine-tuned. 
%For possibility (2), our monotonicity assumptions imply $A(\phi_{z =0} ) \le A(\phi_{\Delta z})$, which goes against self-acceleration.
%Nevertheless this case may offer a loophole to the Compton wavelength theorem. But it also raises other problems. 
%For instance, since $\phi_{\rm in-MW} \le \phi_{z =0} \le \phi_{\Delta z}$ in this case, it is unclear how $\phi$ would dynamically
%settle to its value $\phi_{\rm in-MW}$ deep inside the Milky Way.

Another possibility is to relax the screening condition,
which assumes the existence of {\it test particles}
moving on Jordan frame geodesics. If all astronomical objects used as
dynamical tracers are screened~\cite{Mota:2006ed}, then there is no need
to enforce (\ref{screenedforce}). But this drastic measure
to hide nearly all modified gravity effects leads to a strong backreaction: 
with even the smallest bodies being screened, the effective density sourcing $\phi$ cosmologically
must be a tiny fraction of the total matter density, and
thus the cosmological evolution needs to be reconsidered.

A further possibility is to relax the assumption of a single scalar field.
It is likely possible to extend our no-self-acceleration theorem to a multi-field
version if $V$ and $A$ continue to be monotonic functions
of $\rho$ at equilibrium. It is unclear, however, how the mass bound
would be modified in a multi-field context. Fluctuations around the
effective minimum would be described by a mass matrix,
whose eigenvalues can span a wide range of scales. 
We leave a detailed investigation to future work.

Let us close with an observation on how theories that screen
by the Vainshtein mechanism circumvent our no-go theorems. They replace the potential
$V(\phi)$ by derivative interactions. A key effect is
that the screening condition (\ref{screenedforce}) needs only
hold up to some radius, the so-called Vainshtein radius,
of the object, thus decoupling $\phi_{\rm out}$ from $\phi_{z=0}$.
It would be interesting to investigate whether chameleon-like theories can
also achieve such decoupling.

\smallskip{\em Acknowledgments.}  We thank P.~Brax, C.~van de Bruck, A.-C.~Davis, W.~Hu, A.~Nicolis and A.~Upadhye for helpful discussions. This work is supported in part by
by the DOE grant DE-FG02-92-ER40699 and NASA ATP grant NNX10AN14G (L.H. and J.W.), as well as by NASA ATP grant NNX11AI95G, the Alfred P. Sloan Foundation and NSF CAREER Award PHY-1145525 (J.K.).

%%%%%%%%%%%%%%%%%%%%%%%%%%%%%%%%%%%%%%%%%%%%%%%%%%%%%%%%%%%%%%%%%%%%%%%%%%%%%%%


\begin{thebibliography}{99}
\bibitem{ftnote1}
Relating the Jordan and Einstein frame scale factors by
$a_{\rm J} = A a_{\rm E}$, it is straightforward to show
$[a\ddot a]_{\rm J} - [a \ddot a]_{\rm E} = (A'/A)'$, where
$\dot {}$ denotes derivative with respect to 
(Jordan or Einstein frame) proper time, and ${}'$ denotes
derivative with respect to conformal time.
Thus, if $[a\ddot a]_{\rm E} \le 0$, we must have
$[a\ddot a]_{\rm J} \le (A'/A)'$, implying $1 \lesssim \Delta A/A$
over a (Jordan frame) Hubble time.
\bibitem{ftnote2}
More precisely: $U \cdot \nabla \rho = - \rho \nabla \cdot
U$ for a pressureless fluid.
Here, $\rho = \rho_{\rm E} / A$, $U^\mu$ is the fluid 4-velocity in Einstein
frame ($= A \, \times$ Jordan frame velocity), and indices are contracted with the Einstein frame metric.
\bibitem{ftnote3}
Symmetrons seem to violate this assumption, by
having more than one minimum: different parts of
the universe might inhabit different domains.
This case is nonetheless covered
by our arguments, as long as one interprets the
`field range of interest' as that within one domain.
\bibitem{ftnote4}
For instance, one would like to demand the scalar mass
to be large, or the coupling $\alpha$ to be small, at high $\rho$.
But our theorems do not rely on these additional assumptions.
\end{thebibliography}
\end{document}